\documentstyle[aps,aipbook,psfig]{revtex}

\def\etal{{\it et\ al.}}
\def\simgreat{\mathbin{\lower 3pt\hbox
     {$\rlap{\raise 5pt\hbox{$\char'076$}}\mathchar"7218$}}} 
\def\simless{\mathbin{\lower 3pt\hbox
     {$\rlap{\raise 5pt\hbox{$\char'074$}}\mathchar"7218$}}} 

\begin{document}

\title{Cyclotron Line Formation in a Relativistic Outflow}

\author{Michael Isenberg$^1$, D. Q. Lamb$^1$, and 
John C.L. Wang$^2$
}

\address{$^1$Department of Astronomy and Astrophysics, University of Chicago,
5640 South Ellis Avenue, Chicago, IL  60637 \\
$^2$Department of Astronomy, University of Maryland, College Park, MD  20742-2421}

\maketitle

\begin{abstract}
There is mounting evidence that, if gamma-ray bursters are Galactic in
origin, they are located in a Galactic corona at distances greater
than 100 kpc.  This has created a need to explore new models of
cyclotron line formation.  In most previous calculations the
line-forming region was modeled as a static slab of plasma, optically
thin to continuum scattering, and threaded by a magnetic field of the
order $10^{12}$ gauss oriented normal to the slab.  Such a model is
appropriate, for example, for the magnetic polar cap of a neutron star
with a dipole field.  However, if bursters lie at distances farther
than several hundred parsecs, the burst luminosity exceeds the
magnetic Eddington luminosity, and the plasma in a line-forming region
at the magnetic polar cap would be ejected relativistically along the
field lines.  Mitrofanov and Tsygan have modeled the dynamics of such
an outflow, and Miller {\it et al.}  have calculated the properties of
the cyclotron second and third harmonics, approximating them as due to
cyclotron absorption.  Here we describe Monte Carlo calculations of
cyclotron resonant scattering at the first three harmonics in a
relativistic outflow from the magnetic polar cap, and show that such
scattering can produce narrow lines like those observed by Ginga.
\end{abstract}

\section*{Introduction}

One of the most compelling pieces of evidence that gamma-ray bursts
are Galactic in origin is the observation of absorption-like features
in the spectra of some bursts and the interpretation of these features
as cyclotron lines \cite{Mazets81}.  In particular, the Ginga
observations of three gamma-ray bursts with harmonically spaced lines
\cite{Murakami88} strongly supports this interpretation.  Recent
reports of several highly significant line candidates by the BATSE
\cite{Briggs96a} and Konus \cite{Konus96} groups have further
heightened interest in cyclotron line formation.

Most theoretical models of line formation in gamma-ray bursts assume
physical conditions that are appropriate for burst sources in the
Galactic disk.  For example, in the model of Wang \etal\
\cite{Wang89}, the line-forming region is a static slab of plasma,
optically thin to continuum scattering, and threaded by a uniform
magnetic field $\sim 10^{12}$ gauss oriented along the slab normal.
Such a model is suitable, for example, for the magnetic polar cap of a
neutron star with a dipole field.  Lamb, Wang, and Wasserman [LWW,
\cite{LWW90}] pointed out that if the line-forming region is indeed at
the polar cap, the static model is valid only if the bursters lie at
distances less than several hundred parsecs.  Otherwise, the burst
luminosity is sufficient to create a relativistic plasma outflow along
the field lines.

However, the BATSE burst brightness and sky distributions
\cite{Meegan92,Briggs96b} suggest that if the bursters are Galactic,
they lie in a Galactic corona at distances of 100-400 kpc.  In light
of the BATSE results, it is important to explore line formation models
that are appropriate for sources at these distances.  One possibility
is line formation in a static slab located at the magnetic equator of
a neutron star, where the plasma is magnetically confined near the
surface \cite{Isenberg96,Freeman96}.  In the present work, however, we
explore another possibility: that the lines are formed in a
relativistic outflow.

In an outflow, the variation of the magnetic field and plasma velocity
with altitude tends to broaden the lines.  Miller \etal
\cite{Miller91,Miller92} calculated the properties of the second and
third harmonics, approximating them as due to cyclotron absorption.
They showed that narrow lines can be formed at these harmonics.
Chernenko and Mitrofanov \cite{CM93} calculated the properties of the
first harmonic line, also approximating it as due to absorption, and
found that the formation of a narrow line is possible.  However, such
an approximation is not valid for the first harmonic.  Thus the
question of whether narrow first harmonic {\it scattering} lines can
be formed in an outflow has remained open.  In the present work we use
a Monte Carlo radiative transfer code to calculate the properties of
the first three harmonic lines and address this question.

\section*{Physics of the Line-forming Region}
In our model, photons are injected into the line forming region at a
circular hot spot, with radius $r_{\rm hot}$, located on the surface
of a neutron star and centered on the magnetic pole.  The photons are
distributed uniformly over the hot spot and their directions are
distributed isotropically.  We choose an injected photon number
spectrum that varies inversely with photon energy ($dN_\gamma/dE\
\propto\ E^{-1}$).  The field strength decreases with altitude, z, as
a dipole:

\begin{equation}
B(z) \ = \ B_o\ \left(1 + {z \over R} \right)^{-3},
\end{equation}

\noindent
where $B_o$ is the field strength at the stellar surface, $R$ is the
stellar radius, and $z$ is the altitude above the surface.  Although
the lines of force in a dipole field flare outwards as the altitude
increases, we assume for simplicity that the field lines remain
perpendicular to the surface.  This is a good approximation for $z <<
R$ and $r_{\rm hot} << R$.  Since our Monte Carlo simulations for
$r_{\rm hot}=0.1R$ show that $>90 \%$ of scatters occur at $z<0.1 R$,
we do not expect this assumption to have a significant effect on the
emerging spectrum.

The radiation force accelerates an electron-proton plasma to a flow
velocity, $\beta_f$, which varies with altitude.  Mitrofanov and
Tsygan \cite{MT82} derived the radiation force due to resonant
scattering of an electron located above the center of the hot spot.
At any given altitude, there is an equilibrium velocity at which the
the radiation force on the electron, averaged over the energies and
directions of the photons, is equal to zero.  The equilibrium velocity
is:

\begin{equation}
\beta_e(z) \approx {1 \over 2} \left(1 + {z \over \sqrt{z^2 + r_{\rm
hot}^2}}\right). \label{betae}
\end{equation}

An electron injected at the surface with an initial velocity of zero
accelerates quickly.  From the magnitude of the radiation force, we
can estimate the distance the electron travels before reaching
$\beta_e$.  For a magnetic field strength $B_o\ =\ 1.7 \times 10^{12}$
gauss and an x-ray luminosity between 1 keV and 1 MeV equal to
$10^{40}$ ergs ${\rm s}^{-1}$ (i.e., 1\% of the total burst
luminosity) the distance to $\beta_e$ is $\sim 10^{-7}$ of the stellar
radius.  Once the electron reaches $\beta_e$, its velocity continues
to increase according to eq.(\ref{betae}) until it reaches an altitude
where the radiation becomes too diffuse to provide sufficient energy
for acceleration to continue at the rate required by the equation.  At
this point, the electron's velocity starts to lag behind the
equilibrium velocity.  We estimate that this happens at an altitude
$\sim$ a few stellar radii.  Since most scatterings take place far
below this point, we take $\beta_f = \beta_e$ throughout the
line-forming region.

We emphasize that we calculate $\beta_e$ using the {\it unscattered}
radiation spectrum.  We have not attempted in the present work to
account for the effect on $\beta_e$ of the reduction in photon flux at
the cyclotron energy due to scattering.

We assume that in the frame of reference co-moving with the flow the
distribution of electron velocities is Maxwellian.  LWW showed that
the heating and cooling of the electrons by scattering with the
radiation balances at the Compton equilibrium temperature, $T_c$.
Applying the single-scattering model of LWW to the angular
distribution of radiation in the co-moving frame, we find that $kT_c
\approx \hbar \omega_B/4$, where $\hbar \omega_B$ is the cyclotron
energy.

For burster distances $\sim 100$ kpc, the time scales for energy and
momentum exchange between the electrons and the radiation field are
much shorter than the time scale for establishing a Maxwellian
electron velocity distribution by particle collisions.  The actual
electron distribution is therefore likely to be much narrower than a
Maxwellian, which would narrow the cyclotron lines in the emerging
spectrum.  Thus our assumption of a Maxwellian electron velocity
distribution in the present calculation is conservative.

Following Miller \etal \cite{Miller91,Miller92}, we calculate the
density of the plasma as a function of altitude from the continuity
equations for conservation of mass and magnetic flux:

\begin{equation}
n_e(z) = n_{e,o} \ {B(z) \over B(0)} \ {\beta_f(0) \over \beta_f(z)}
\end{equation}
\noindent
where $n_{e,o}$ is the plasma density at the stellar surface.
\vfill\eject

\begin{figure}
\centerline{ \psfig{file=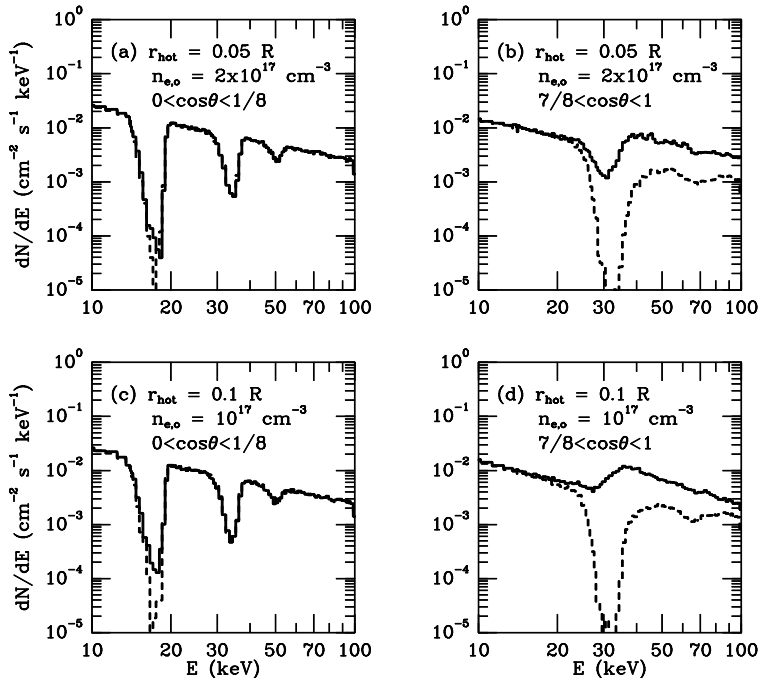,width=10cm,angle=-90}}
\caption{ Theoretical photon number spectra for $B_o = 1.7 \times
10^{12}$ gauss and electron column depth $n_{e,o} r_{\rm hot} =
10^{22} {\rm cm}^{-2}$.  Resonant scattering (solid line) and pure
absorption (dashed line) spectra are shown for two viewing angles,
$\theta$ with respect to the magnetic field.  Top panels: hot spot
radius, $r_{\rm hot} = 0.05 R$, where $R$ is the stellar radius;
bottom panels: $r_{\rm hot} = 0.1 R$.  Narrow absorption-like lines
occur for most viewing angles, while broad absorption-like or
emission-like features occur when the viewing angle lies along the
field.}
\label{spectra}
\end{figure}

\section*{Results and Discussion}
We calculate the emerging radiation spectrum using a Monte Carlo
radiative transfer code similar to the one described by Wang \etal
\cite{Wang89}.  The code is valid for line forming regions where the
cyclotron first harmonic is optically thick in the line core but
optically thin in the wings.  The cross sections are summed over final
and averaged over initial polarizations.  We use exact relativistic
kinematics and zero natural line width.  We include scattering at the
first three harmonics and photon spawning.

The emerging spectrum of radiation is shown at two viewing angles in
Figure 1.  $r_{\rm hot}=0.05 R$ in the top panels and $0.1 R$ in the
bottom panels.  In each case $B_o \ = \ 1.7 \times 10^{12}$ gauss and
the electron column depth $n_{e,o} r_{\rm hot} = 10^{22} {\rm
cm}^{-2}$.

The behavior of the spectra is explained by the high velocity of the
plasma, which causes scattered photons to be beamed along the field.
Consequently, when the spectra are viewed perpendicular to the field
(left panels) the scattered spectra are almost identical to pure
absorption spectra.  Although we expect this in the second and third
harmonics, it is also the case for the first harmonic.  In both
spectra, the equivalent widthes in the first and second harmonics are
$W_{E1}\approx 4.7$ and $W_{E2} \approx 6.2$ keV.  By comparison, in
GB880205, observed by Ginga, $W_{E1} = 3.7$ and $W_{E2}=9.1$ keV
\cite{Fenimore88}.  The narrowness of the lines is due to the finite
radius of the hot spot.  Photons redward of the line are normally
capable of scattering at high altitudes where the cyclotron energy is
smaller, thus broadening the line.  However, a photon moving at a
large angle to the field escapes through the sides of the cylinder of
outflowing plasma before reaching the altitude where it would scatter.

The beaming of scattered photons also accounts for the properties of
the lines when viewed along the field (right panels).  In Figure 1b,
the first harmonic scattering line has been almost entirely filled in,
compared with the first harmonic absorption line.  Only a shallow line
remains.  When the hot spot radius is larger (Figure 1d), photons
scatter at higher altitudes where the magnetic field strength is
smaller.  Consequently, scattered photons emerge at lower energies and
fill in the absorption line entirely, forming a broad emission-like
feature.

Our calculations suggest that a relativistic outflow is able to form
cyclotron scattering lines with properties similar to the lines
observed by Ginga, provided the hot spot is a small fraction of the
stellar surface.  In the future we propose to confirm this suggestion
by more detailed calculations and a fit of the model spectra to the
Ginga observations.  If confirmed, our results would imply that the
interpretation of the observed features as cyclotron lines does not
rule out burst sources in a Galactic corona.

\end{document}